\documentclass[pra,twocolumn,floatfix,a4paper,superscriptaddress,article]{revtex4}
\usepackage{amssymb}
\usepackage{amsfonts}

\usepackage{bm,color,amsmath,txfonts}
\usepackage{graphicx}
\usepackage{siunitx}
\usepackage{subfigure}
\usepackage{verbatim}
\usepackage{dcolumn}
\usepackage{bm}
\usepackage{epsf}
\usepackage{xcolor}
\usepackage{hyperref}
\usepackage{hhline}
\usepackage{float}
\usepackage{enumerate}
\usepackage{bbm}
\usepackage{lipsum}
\usepackage{mathrsfs}
\begin{document}
\title[Short Title]{Strong squeezing and perfect one-way EPR steering in electro-optomechanical system}
\author{Qing-Min Zeng}
\affiliation{State Key Laboratory of Quantum Optics Technologies and Devices, Institute of Opto-Electronics, College of Physics and Electronic Engineering, Shanxi University, Taiyuan, Shanxi 030006, China}
\affiliation{Collaborative Innovation Center of Extreme Optics, Shanxi University, Taiyuan 030006, China}
\author{A-Peng Liu}
\affiliation{State Key Laboratory of Quantum Optics Technologies and Devices, Institute of Opto-Electronics, College of Physics and Electronic Engineering, Shanxi University, Taiyuan, Shanxi 030006, China}
\affiliation{Collaborative Innovation Center of Extreme Optics, Shanxi University, Taiyuan 030006, China}
\affiliation{Shanxi Institute of Technology, Yangquan, Shanxi 045000, China}
\author{Qi Guo\footnote{E-mail: qguo@sxu.edu.cn}}
\affiliation{State Key Laboratory of Quantum Optics Technologies and Devices, Institute of Opto-Electronics, College of Physics and Electronic Engineering, Shanxi University, Taiyuan, Shanxi 030006, China}
\affiliation{Collaborative Innovation Center of Extreme Optics, Shanxi University, Taiyuan 030006, China}

\begin{abstract}

We consider a three-mode electro-optomechanical system in which a mechanical oscillator is coupled to an optical cavity and a LC circuit through radiation pressure and capacitive coupling, respectively. By controlling the two-tone driving of the optical cavity and the microwave driving of the LC circuit, the strong squeezing of both the microwave field of the LC circuit and the mechanical mode will be obtained. Moreover, by further altering the driving power of the system, the perfect one-way Einstein-Podolsky-Rosen (EPR) steering between the optical cavity and the mechanical oscillator will be generated. The degree of the one-way EPR steering can be controlled by the driving light of the system. We also show the robustness of the squeezing and EPR steering against environmental temperature. This scheme may provide a promising platform for quantum information processing and microwave quantum communication.

\end{abstract}
\maketitle

\section{Introduction}

Over the past few decades, quantum mechanics has developed rapidly, gradually evolving from theoretical concept research to experimentally feasible practical applications~\cite{1}. Meanwhile, significant progresses have also been made in cavity optomechanics and circuit quantum electrodynamics. The former allows the optical field in the optical cavity to be coupled to mechanical displacements through radiation pressure~\cite{2}, and can be further directly or indirectly coupled to other subsystems in other ways, such as another optical cavity~\cite{3,4,5}, atomic ensemble~\cite{6,7,8}, etc. The latter involves quantization treatment of nonlinear superconducting circuits and investigation of their interaction with microwave fields~\cite{9,10,62,11}, which is of great significance in the fields of quantum communication and quantum computation. One of the most significant achievements in these fields is the coupling and transduction between optical cavities and quantum circuits and macroscopic mechanical oscillators~\cite{12,13,14,15,16}, which can be used to study locality based on quantum measurement and quantum information protocols~\cite{17,18}. Therefore, it is naturally to consider for finding a way to achieve the mutual transduction of quantum states
in these two different frequency bands by
using macroscopic mechanical oscillators as intermediaries~\cite{19,20}. Such a system that can serve as a quantum interface is called an electro-optomechanical system and can realize the storage and redistribution of quantum information.
 
Theories and experiments have shown that an electro-optomechanical system formed by coherent coupling among optical photons, mechanical mode and microwave photons can be used as an emerging promising platform for preparing entangled states of microwave photons, optical photons and mechanical mode~\cite{19,20,21,22,23,24,25}. Moreover, the cooling of the ground states of mechanical oscillators~\cite{26,27} and quantum steering~\cite{28,29} can be achieved through this system. In the system, the conductive nanomechanical membrane serves as a part of the capacitance of the LC circuit. When the optical cavity is driven by a laser, the mechanical displacement caused by the radiation pressure can change the total capacitance~\cite{19}, enabling the microwave field of the LC circuit couple to the mechanical displacement and further couple to the electromagnetic field of the optical cavity. Thus, a connection between two different frequency bands is achieved and experiments have also proved that this system can realize the transduction between such two frequency bands~\cite{30,31,32,33,34}. Due to the fact that the modes of different frequency bands can interact with each other, the electro-optomechanical system is of great significance in the preparation of nonclassical states of microwaves and microwave quantum communication.

So far, there are severals methods for preparing squeezed states of microwave fields or achieveing EPR steering in electro-optomechanical systems. For example, introducing superconducing charge qubits into the system under strong dispersion limit, which enhances the nonlinear effect between the microwave field and mechanical displacement, thereby enabling microwave field squeezing~\cite{10}. Other approaches involve dual-light-driven and modulated laser-driven optical cavities within electro-optomechanical systems~\cite{35,36}. However, the system described in Ref.~\cite{35} is somewhat complex. Here, we do not modify the original electro-optic mechanical system's modules, instead, by using dual-light-driven optical cavities with different frequencies and amplitudes, we prove that this mechanical squeezing can be transferred to the microwave mode of the LC circuit. Through further adjusting the driving fields of the system, we can achieve perfect one-way EPR steering between the optical mode and mechanical mode. To our knowlodge, there is no electro-optomechanical system capable of achieving these non-classical states simultaneously. We have noticed in Ref.~\cite{19} that red-blue detuned driven optical cavities and LC circuits were adopted, respectively, and generates microwave-optical entanglement. Here, if merely the dual-light-driven optical cavities are introduced, although mechanical oscillator can be cooled to the quantum ground state and generate optical mechanical entangled state, the system is unable to generate the squeezed state of the microwave field. In order to transfer the squeezing from the mechanical mode to the microwave mode, we made the LC circuit driven by red detuning. The results show that the strong squeezing state of microwave field and mechanical mode can be prepared in this system. Quantum squeezing is of great significance in quantum information processing~\cite{37,38,39,40,41} and improving the accuracy of quantum measurements~\cite{42,43,44}. Quantum steering, as a strict subset of entanglement, especially the one-way quantum steering that has been theoretically~\cite{46,47,48,49,50} and experimentally~\cite{51,52,53,54,55,56} realized at present, has significant application value for quantum secure communication.

The organizational structure of this article is as follows: In Sec.$~$\ref{Sec.II}, we present the model diagram of the electro-optomechanical system, giving the Hamiltonian of the system and deriving the quantum Langevin equations. Subsequently, we obtain the steady-state solution of the dynamical equations and the effective Hamiltonian. In Sec.$~$\ref{Sec.III}, we demonstrated the steady-state squeezing of mechanical mode and microwave field. In Sec.$~$\ref{Sec.IV}, we adopt logarithmic negativity to quantize entanglement and proof the effect of driving optical power on the magnitude of one-way EPR steering. Finally, in Sec.$~$\ref{Sec.V}, we provide a summary.

\section{The model and the Quantum Langevin Equations}\label{Sec.II}

\begin{figure}[htb]
	\centering
	\includegraphics[width=\linewidth]{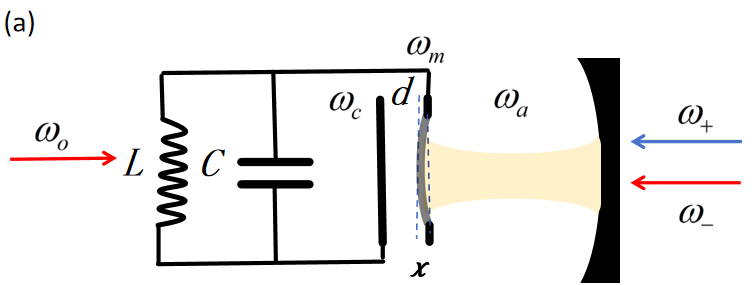}
	\includegraphics[height=2.0cm,width=3.0cm]{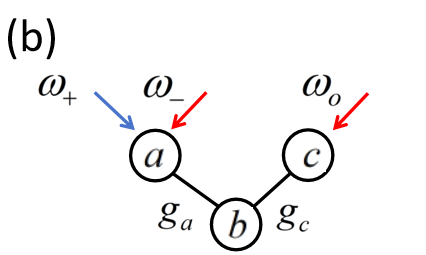}
	\includegraphics[height=2.0cm,width=5.3cm]{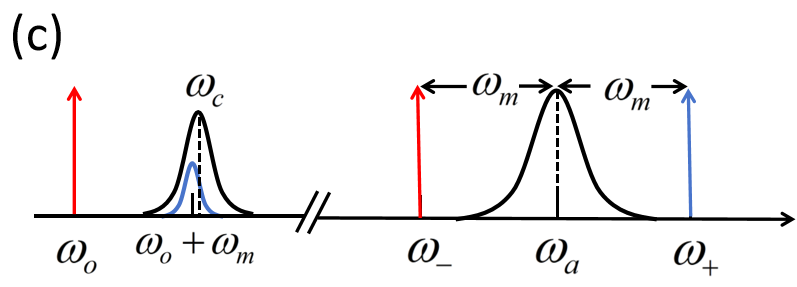}
\caption{(a) Schematic diagram for the electro-optomechanical system we are studying: a membrane with an optical coating, a capacitor and two inductors form an LC circuit, which acts as a microwave cavity. This microwave cavity is coupled to a mechanical resonator via the capacitance change induced by radiation pressure. Additionally, it is indirectly coupled to an optical cavity which is composed of an input mirror and the mechanical resonator. (b) The optical cavity mode $\hat{a}$, driven by two different frequencies of lasers ($\omega_{\pm}$), and the microwave mode $\hat{c}$, driven by a microwave of frequency $\omega_{o}$, are both nonlinearly coupled to the mechanical mode $\hat{b}$. (c) The frequency relationship of the system: the frequency difference between the two driving frequencies of the driving cavity mode $\hat{a}$ is twice the mechanical mode frequency $\omega_{m}$. Due to the scattering of mechanical motion, sideband $(\omega_o + \omega_m)$ is generated, and microwave mode $\hat{c}$ resonates with the anti-Stokes sideband.}
\label{1}
\end{figure}

The electro-optomechanical system proposed is shown in Fig.$~$\ref{1}. It consists of three modules: an optical cavity (OC), a mechanical resonator (MR) and an LC circuit that can be regarded as a microwave cavity (MC). The MR is coupled on one side to the OC and on the other side capacitively to the MC. Adding an optical coating on the capacitor plates that are coupled with the OC, motion function of the electrode plate is similar to that of the Fabry-Perot optical cavity's micromirror~\cite{19}. Let the driving frequencies of the dual-light laser field be $\omega_{+}$ and $\omega_{-}$, and the driving frequency of the microwave cavity be $\omega_{o}$. Then the Hamiltonian of the system is obtained as~\cite{19,57,58}
\begin{align}
	&\hat{H}=\omega_{a}\hat{a}^\dagger\hat{a}+\frac{{p}^2}{2m}+ \frac{1}{2}m\omega_m^2x+\frac{Q^2}{2[C+C_0 (x)]}+\frac{\Phi^2}{2L}+\notag\\
	&\quad~~~~\hbar{G}_{a}\hat{a}^\dagger\hat{a}x+i\hbar[(E_{+}e^{-i\omega_+t}+E_{-}e^{-i\omega_-t})\hat{a}^\dagger-H.c.]-e(t)Q,
	\label{x1}
\end{align}
here, $\hat{a}~(\hat{a}^\dagger)$ denotes the annihilation (creation) operator of the OC  $([\hat{a},\hat{a}^\dagger] = 1)$ with frequency $\omega_a$, $x$ and $p$ respectively denote the standard position and momentum operator of the MR with frequency $\omega_m$, while $L$ and $C$ respectively represent the equivalent inductance and capacitance of the MC. $\Phi$ and $Q$ respectively indicate the magnetic flux passing through the equivalent inductance and the charge quantity of the equivalent capacitance. The first term of $\hat H$ represents the free Hamiltonian of the OC, the second and third terms represent the free Hamiltonian of the MR, the fourth and fifth terms represent the free Hamiltonian of the MC and its coupling with the MR, the sixth term represents the coupling between the OC and the MR, $G_a = \frac{\omega_a}{\zeta}\sqrt{\frac{\hbar}{m\omega_m}}$ is the single photon-optomechanical coupling strength, $\zeta$ represents the cavity length of the OC, and the last two terms respectively represent the dual-light driving term of the OC and the driving term of the microwave field. The two driving fields of the OC are respectively located in the Stokes scattering sideband $(\omega_- = \omega_a - \omega_m)$ and the anti-Stokes scattering sideband $(\omega_+ = \omega_a + \omega_m)$,  and $E_\pm = \sqrt{2P_\pm\kappa_a/\hbar\omega_\pm}$, $e(t) = -i\sqrt{2\hbar\omega_cL}E_c(e^{i\omega_ot}-e^{-i\omega_ot})$. The coupling between the MC and the MR is due to the mechanical displacement caused by radiation pressure. This displacement leads to a change in the capacitance of LC circuit, and this change has a functional relationship with the mechanical displacement, that is $C_0(x)$,

\noindent and $C_0(x) = C_0[1 + x(t)/d]$, $d$ represents the gap between the electrode plate and the membrane, and $C_0$ represents the capacitance at the equilibrium position. Since $x$ is very small, for the total capacitance of the LC circuit, we take the first-order term in the Taylor expansion:
\begin{align}
	\frac{Q^2}{2[C+C_0 (x)]} =  \frac{Q^2}{2C_{s}} - \frac{\epsilon}{2dC_{s}}x(t)Q^{2},
	\label{x2}
\end{align}
where $C_s = C+C_0$, $\epsilon = C_0/C_s$. It should be more convenient to describe the hamiltonian of the MC (MR) in $\hat{H}$ with its creation operator $\hat{c}^\dagger$ ($\hat{b}^\dagger$) and annihilation operator $\hat{c}$ ($\hat{b}$), $([\hat{c},\hat{c}^\dagger] = 1,[\hat{b},\hat{b}^\dagger] = 1)$. The total Hamiltonian then can rewritten as
\begin{align}
	&\hat{H}=\hbar\omega_{a}\hat{a}^\dagger\hat{a}+\hbar\omega_{m}\hat{b}^\dagger\hat{b}+\hbar\omega_{c}\hat{c}^\dagger\hat{c}+\hbar{g}_{a}\hat{a}^\dagger\hat{a}(\hat{b}^\dagger+\hat{b})\notag\\
	&\quad~~~~\hbar{g}_{c}\hat{c}^\dagger\hat{c}(\hat{b}^\dagger+\hat{b})+i\hbar[(E_{+}e^{-i\omega_+t}+E_{-}e^{-i\omega_-t})\hat{a}^\dagger-H.c.]\notag\\
	&\quad~~~-i\hbar E_c(\hat{c}^\dagger+\hat{c})(e^{i\omega_ot}-e^{-i\omega_ot}),
	\label{x3}
\end{align}
where
\begin{align}
	&\quad\quad\hat{b}=\sqrt{\frac{m\omega_m}{2\hbar}}\hat{x}+\frac{i}{\sqrt{2\hbar m\omega_m}}\hat{p},\notag\\
	&\quad\quad\hat{c}=\sqrt{\frac{L\omega_c}{2\hbar}}\hat{Q}+\frac{i}{\sqrt{2\hbar L\omega_c}}\hat{\Phi},\notag\\
	&\quad g_a = \frac{\omega_a}{\zeta}\sqrt{\frac{\hbar}{2m\omega_m}}, g_c = \frac{\epsilon\omega_c}{2d}\sqrt{\frac{\hbar}{2m\omega_m}},
	\label{x4}
\end{align}
here, $\omega_c$ is the resonant frequency of the MC, and its magnitude is $1/\sqrt{LC_s}$.
Transform $\hat{H}$ in a to the interaction picture with respect to $\hat{H_0} = \hbar\omega_o\hat{b}^\dagger\hat{b}$, ignoring the high-frequency oscillation terms $\pm2\omega_o$, and then return to the original picture. At this time, the Hamiltonian can be written as
\begin{align}
	&\hat{H}=\hbar\omega_{a}\hat{a}^\dagger\hat{a}+\hbar\omega_{m}\hat{b}^\dagger\hat{b}+\hbar\omega_{c}\hat{c}^\dagger\hat{c}+\hbar{g}_{a}\hat{a}^\dagger\hat{a}(\hat{b}^\dagger+\hat{b})\notag\\
	&\quad~~~~\hbar{g}_{c}\hat{c}^\dagger\hat{c}(\hat{b}^\dagger+\hat{b})+i\hbar[(E_{+}e^{-i\omega_+t}+E_{-}e^{-i\omega_-t})\hat{a}^\dagger-H.c.]\notag\\
	&\quad~~~+i\hbar E_c(\hat{c}^\dagger e^{-i\omega_ot}-\hat{c}e^{i\omega_ot}).
	\label{x5}
\end{align}

 Next, the quantum Langevin equations are employed to describe the dynamics of the system. These equations incorporate the photon losses of the two cavities and the damp of the mechanical resonator, as well as their respective input noises. The resulting nonlinear quantum Langevin equations are
\begin{align}\label{x6}
	&\dot{\hat{a}} = \left( -i\omega_a - \kappa_a \right) \hat{a} - ig_a \hat{a} \left( \hat{b}^\dagger + \hat{b} \right) + E_+ e^{-i\omega_{+}t} + E_- e^{-i\omega_{-}t} \notag\\
	&\quad+ \sqrt{2\kappa_a} \hat{a}^{\text{in}},\notag\\
	&\dot{\hat{b}} = \left( -i\omega_m - \gamma_m \right) \hat{b} - ig_a \hat{a}^\dagger \hat{a} - ig_c \hat{c}^\dagger \hat{c} + \sqrt{2\gamma_m} \hat{b}^{\text{in}},\notag\\
	&\dot{\hat{c}} = \left( -i\omega_c - \kappa_c \right) \hat{c} - ig_c \hat{c} \left( \hat{b}^\dagger + \hat{b} \right) + E_c e^{-i\omega_o t} + \sqrt{2\kappa_c} \hat{c}^{\text{in}},
\end{align}
here, $\kappa_a$ and $\kappa_c$ represent the dissipation of the OC and the MC respectively, while $\gamma_m$ is the damping rate of the MR. $\hat{a}^{\text{in}}$ ($\hat{b}^{\text{in}}/ \hat{c}^{\text{in}}$) is the input noise operator of the OC (MR/MC). Under the Markov approximation, their zero-mean correlation functions are~\cite{57,58}
\begin{align}
	&\langle \hat{a}^{\mathrm{in}}(t)\hat{a}^{\mathrm{in},\dagger }(t') \rangle=[N_a(\omega_a)+1]\delta(t-t'),\notag\\
	&\langle \hat{a}^{\mathrm{in},\dagger }(t)\hat{a}^{\mathrm{in}}(t') \rangle=N_a(\omega_a)\delta(t-t'),\notag\\
	&\langle \hat{b}^{\mathrm{in}}(t)\hat{b}^{\mathrm{in},\dagger }(t') \rangle=[N_b(\omega_b)+1]\delta(t-t'),\notag\\
	&\langle \hat{b}^{\mathrm{in},\dagger }(t)\hat{b}^{\mathrm{in}}(t') \rangle=N_b(\omega_b)\delta(t-t'),\notag\\
	&\langle \hat{c}^{\mathrm{in}}(t)\hat{c}^{\mathrm{in},\dagger }(t') \rangle=[N_c(\omega_c)+1]\delta(t-t'),\notag\\
	&\langle \hat{c}^{\mathrm{in},\dagger }(t)\hat{c}^{\mathrm{in}}(t') \rangle=N_c(\omega_c)\delta(t-t'),
	\label{x7}
\end{align}
where $N_c(\omega_c)=[e^{\hbar \omega_c/k_BT}-1]^{-1}$ and $N_b(\omega_b)=[e^{\hbar \omega_b/k_BT}-1]^{-1}$ respectively represent the average number of thermal photon of microwave mode and the average thermal excitation number of mechanical vibrations of the MR. In the optical frequency regime, $\frac{\hbar \omega_a}{k_BT}\gg1$, so $N_a(\omega_a)\approx0$, thus the thermal photon number of the optical mode can be neglected. $T$ represents the ambient temperature, and $k_B$ is the Boltzmann constant. In the system, since the three driving lights we employ are strong, linearization processing can be carried out. We can write the operators of each mode in this form: $\hat{a}=a_s+\delta\hat{a}$ , $\hat{b}=b_s+\delta\hat {b}$ , $\hat{c}=c_s+\delta\hat {c}$. After substituting these expressions into Eq.$~$(\ref{x6}), the steady-state averages and the dynamics of quantum fluctuations for each mode can be derived. Since we use two lasers with different frequencies to drive the optical cavity, the steady-state mean value of the optical mode is expressed as $a_s = a_+e^{-i\omega_+t}+a_-e^{-i\omega_-t}$, while the expression of the steady-state mean value of the microwave mode is $c_s = a_ce^{-i\omega_ot}$. The optically induced frequency shifts $g_a(b_s^{*}+b_s)$ and $g_c(b_s^{*}+b_s)$ caused by radiation pressure are significantly smaller than the intrinsic mechanical frequency $\omega_m$. Therefore, we can safely neglect these minor frequency shifts when calculating the steady-state mean values. In this configuration, we set the microwave cavity to be red-detuned, $\omega_c-\omega_o=\omega_m$, then we can calculate that the steady-state averages are
\begin{align}
	&a_\pm=\frac{E_\pm}{i(\omega_a-\omega_\pm)+\kappa_a}=\frac{E_\pm}{\mp i\omega_m+\kappa_a},\notag\\
	&\quad\quad\quad\quad~~ a_c=\frac{E_c}{i\omega_m+\kappa_c},
	\label{x8}
\end{align}
by inserting $a_s$ and $c_s$ into Eq.$~$(\ref{x5}) while keeping only operator terms up to second order, and neglecting the small frequency shifts, we derive the effective Hamiltonian representation in the rotating frame of $H_1 = \omega_{a}\hat{a}^\dagger\hat{a}+\omega_{m}\hat{b}^\dagger\hat{b}+\omega_{c}\hat{c}^\dagger\hat{c}$ ($\hbar = 1$):
\begin{align}
	\hat{H}_{\text{eff}} &= G_+ \left( \delta\hat{a}\delta\hat{b} + \delta\hat{a}^{\dagger}\delta\hat{b}^{\dagger} \right) 
	 + G_- \left( \delta\hat{a}\delta\hat{b}^{\dagger} + \delta\hat{a}^{\dagger}\delta\hat{b} \right) \notag \\
	&\quad + G_{c} \left( \delta\hat{c}\delta\hat{b}^{\dagger} + \delta\hat{c}^{\dagger}\delta\hat{b} \right),
	\label{x9}
\end{align}
where $G_+ = g_a a_+$, $G_- = g_a a_-$, $G_c = g_c a_c$. Taking into account the dissipation and input noise of the three modes, we rewrite the quantum fluctuation dynamics as:
\begin{align}\label{x10}
	\delta\dot{\hat{a}} &= -\kappa_a \delta\hat{a} - iG_+ \delta\hat{b}^\dagger - iG_- \delta\hat{b} + \sqrt{2\kappa_a} \hat{a}^{\text{in}}, \notag\\
	\delta\dot{\hat{b}} &= -\gamma_m \delta\hat{b} - iG_+ \delta\hat{a}^\dagger - iG_- \delta\hat{a} - iG_c \delta\hat{c} + \sqrt{2\gamma_m} \hat{b}^{\text{in}}, \notag\\
	\delta\dot{\hat{c}} &= -\kappa_c \delta\hat{c} - iG_c \delta\hat{b} + \sqrt{2\kappa_c} \hat{c}^{\text{in}},
\end{align}
directly solving for Eq.$~$(\ref{x10}) is computationally challenging, to simplify the calculation, we orthogonalize the fluctuation operators in Eq.$~$(\ref{x10}) and express them in the following form ($j = a, b, c$):
\begin{align}\label{x11}
	\delta{\hat{X}}_{j}=\dfrac{1}{\sqrt{2}}(\delta{\hat{j}}+\delta{\hat{j}}^\dagger), ~~\delta{\hat{Y}}_{j}=\dfrac{1}{i\sqrt{2}}(\delta{\hat{j}}-\delta{\hat{j}}^\dagger),
\end{align}
and the corresponding noise quadrature operators:
\begin{align}\label{x12}
	{\hat{X}}_{j}^{in}=\dfrac{1}{\sqrt{2}}({\hat{j}}^{in}+{\hat{j}}^{in\dagger}), ~~{\hat{Y}}_{j}^{in}=\dfrac{1}{i\sqrt{2}}({\hat{j}}^{in}-{\hat{j}}^{in\dagger}),
\end{align}
and Eq.$~$(\ref{x10}) become:
\begin{align}\label{x13}
	\delta\dot{\hat{X}}_a &= -\kappa_a \delta \hat{X}_a + (G_- - G_+) \delta \hat{Y_b} + \sqrt{2\kappa_a} \hat{X}_a^{\text{in}}, \notag\\
	\delta\dot{\hat{Y}}_a &= -\kappa_a \delta \hat{Y}_a - (G_- + G_+) \delta \hat{X}_b + \sqrt{2\kappa_a} \hat{Y}_a^{\text{in}}, \notag\\
	\delta\dot{\hat{X}}_b &= (G_- - G_+) \delta \hat{Y}_a - \gamma_m \delta \hat{X}_b + G_c \delta \hat{Y}_c + \sqrt{2\gamma_m} \hat{X}_b^{\text{in}}, \notag\\
	\delta\dot{\hat{Y}}_b &= -(G_- + G_+) \delta \hat{X}_a - \gamma_m \delta \hat{Y}_b - G_c \delta \hat{X}_c + \sqrt{2\gamma_m} \hat{Y}_b^{\text{in}}, \notag\\
	\delta\dot{\hat{X}}_c &= G_c \delta \hat{Y}_b - \kappa_c \delta \hat{X}_c + \sqrt{2\kappa_c} \hat{X}_c^{\text{in}}, \notag\\
	\delta\dot{\hat{Y}}_c &= -G_c \delta \hat{X}_b - \kappa_c \delta \hat{Y}_c + \sqrt{2\kappa_c} \hat{Y}_c^{\text{in}}.
\end{align}
Eq.$~$(\ref{x13}) can also be expressed in matrix form as:
\begin{align}\label{x14}
	\dot{u}(t)=A(t)u(t)+n(t),
\end{align}
where, $u^T(t)=(\delta{\hat{X}}_{a},\delta{\hat{Y}}_{a},\delta{\hat{X}}_{b},\delta{\hat{Y}}_{b},\delta{\hat{X}}_{c},\delta{\hat{Y}}_{c})$, and $n^T(t)=(\sqrt{2\kappa_a}\hat{X}_{a}^{in},\sqrt{2\kappa_a}{\hat{Y}}_{a}^{in},\sqrt{2\gamma_m}\hat{X}_{b}^{in},\sqrt{2\gamma_m}{\hat{Y}}_{b}^{in},\sqrt{2\kappa_c}\hat{X}_{c}^{in},\sqrt{2\kappa_c}{\hat{Y}}_{c}^{in})$, and
\begin{equation}
	\mathbf{A} = 
	\resizebox{0.8\linewidth}{!}{$
	\begin{pmatrix}
	\begin{smallmatrix}
		-\kappa_a & 0 & 0 & (G_- - G_+) & 0 & 0 \\
		0 & -\kappa_a & -(G_- + G_+) & 0 & 0 & 0 \\
		0 & (G_- - G_+) & -\gamma_m & 0 & 0 & G_c \\
		-(G_- + G_+) & 0 & 0 & -\gamma_m & -G_c & 0 \\
		0 & 0 & 0 & G_c & -\kappa_c & 0 \\
		0 & 0 & -G_c & 0 & 0 & -\kappa_c		 
	\end{smallmatrix}
	\end{pmatrix}
	$}.
	\label{x15}
\end{equation}

Given that all input noise fields are in Gaussian states and the system's dynamics remain linearized, the system can be described by a $6\times6$ covariance matrix (CM) represented by $V$, the CM can be obtained by solving the Lyapunov equation~\cite{57}:
\begin{align}\label{x16}
	A(t)V+VA(t)^T=-D,
\end{align}
here, its form is not presented because its explicit solution is overly complicated, where the diffusion matrix D is defined by $\delta(t-t') D_{kl}=\langle n_k(t)n_l(t')+n_l(t')n_k(t)$, yielding the final result: $D=\mathrm{diag}[\kappa_a,\kappa_a,\gamma_{m}(2N_m+1),\gamma_{m}(2N_m+1),\kappa_{c}(2N_c+1),\kappa_{c}(2N_c+1)]$. If the real part of the maximum eigenvalue of matrix $A$ is less than 0, the system is stable.

\section{Steady-state quantum squeezing of mechanical and microwave modes}\label{Sec.III}

To investigate the squeezing generation mechanism, we introduce two Bogoliubov modes corresponding to the mechanical ($\beta_1$) and microwave ($\beta_2$) mode respectively:
\begin{align}
	\begin{cases} 
		\beta_1 &= \delta \hat{b} \cosh r + \delta \hat{b}^\dagger \sinh r \\ 
		\beta_2 &= \delta \hat{c} \cosh r + \delta \hat{c}^\dagger \sinh r
	\end{cases},
	\label{x17}
\end{align}
where $r$ is the squeezing parameter defined as $tanhr = sinhr/coshr=G_+/G_-$, $coshr = G_-/\sqrt{G_-^2-G_+^2}$ and $sinhr = G_+/\sqrt{G_-^2-G_+^2}$. By substituting Eq.$~$(\ref{x17}) into Eq.$~$(\ref{x9}), the effective Hamiltonian takes the revised form:
\begin{equation}
	\begin{aligned}
		H'_{\text{eff}} &= \frac{G_c}{G'^2} \Big[ \left( G_-^2 + G_+^2 \right) \left( \beta_2 \beta_1^\dagger + \beta_2^\dagger \beta_1 \right) \\
		&\quad - 2G_-G_+ \left( \beta_2 \beta_1 + \beta_2^\dagger \beta_1^\dagger \right) \Big] \\
		&\quad + G' \left( \delta \hat{a} \beta_1^\dagger + \delta \hat{a}^\dagger \beta_1 \right),
		\label{x18}
	\end{aligned}
\end{equation}
here $G'=\sqrt{G_-^2-G_+^2}$ represents the effective coupling strength between the optical mode and Bogoliubov mode of the MR. The third term in Eq.$~$(\ref{x18}) represents a beam-splitter-type interaction that can effectively cool the Bogoliubov mode of the MR to its quantum ground state, the squeezed state of the mechanical mode is thereby generated. From the first and second terms in Eq.$~$(\ref{x18}), we observe that the mechanical and microwave mode exhibit both beam-splitter-type interactions (enabling squeezing transfer) and parametric amplification-type interactions. Therefore, under dual-light driving, it is possible to simultaneously generate the squeezed state of both mechanical mode and microwave mode.

\begin{figure}
	\centering
	\includegraphics[height=3.5cm,width=4.2cm]{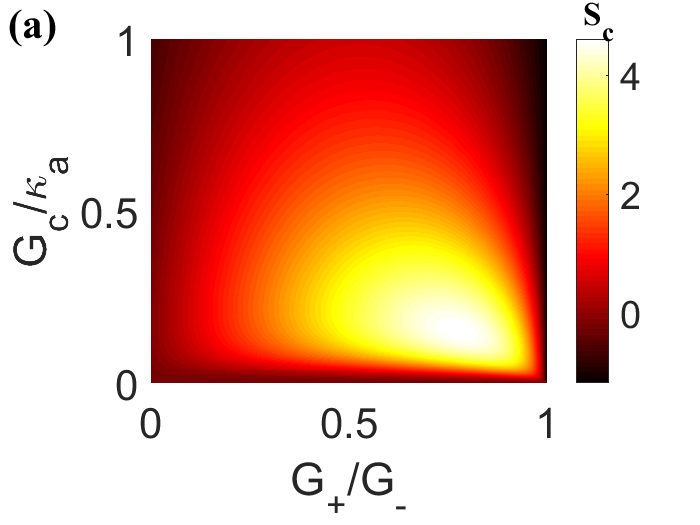}
	\includegraphics[height=3.5cm,width=4.2cm]{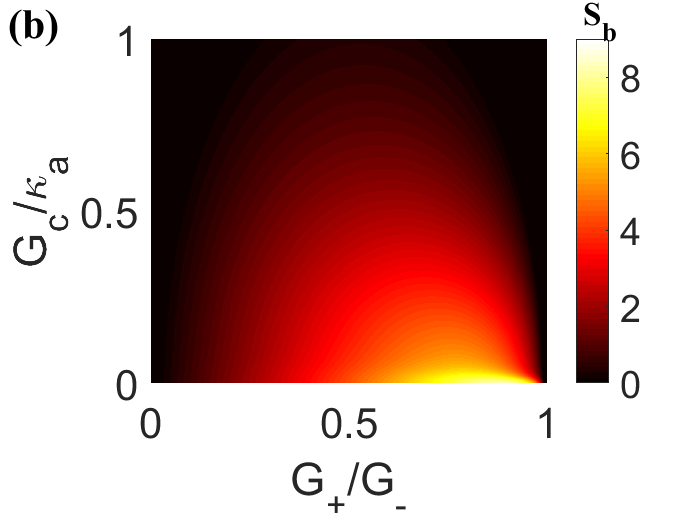}
	\includegraphics[height=3.5cm,width=4.2cm]{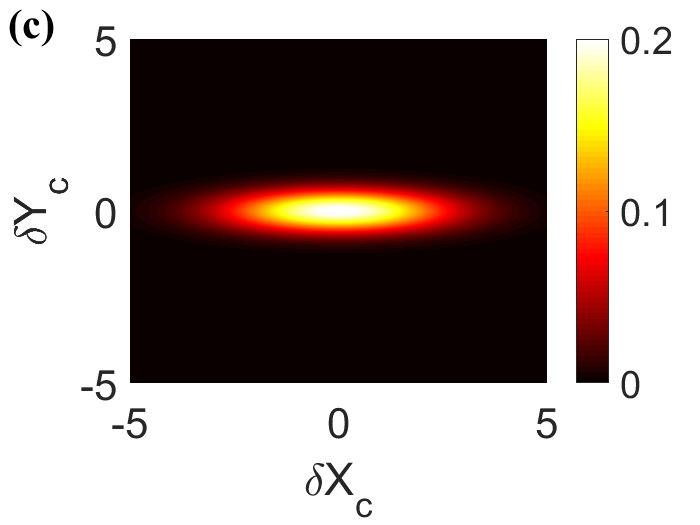}
	\includegraphics[height=3.5cm,width=4.2cm]{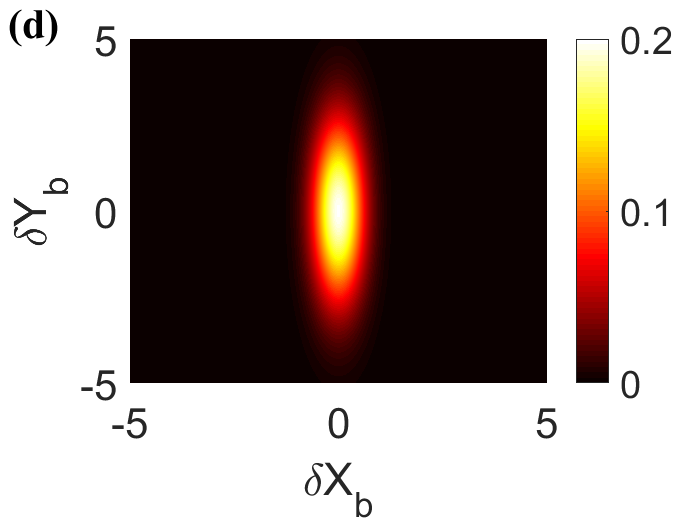}
	\caption{(a) The squeezing degree of the microwave mode ($\hat{c}$) and mechanical mode ($\hat{b}$) versus parameter $G_+/G_-$ and $G_c/\kappa_a$. The Wigner function of the microwave mode (c) and the mechanical mode (d) with $G_-=\kappa_a$, $G_+=0.78G_-$.}
	\label{2}
\end{figure}

The expression for steady-state squeezing (in dB) is defined as $S=-10\log_{10}{\left\langle \delta Q^{2}\right\rangle }/{\left\langle \delta Q^{2}\right\rangle }_{vac} $ with ${\left\langle \delta Q^{2}\right\rangle }_{vac}=0.5$ refers to the vacuum fluctuation, and ${\left\langle \delta Q^{2}\right\rangle }$ represents the orthogonal components of the three modes. In fig.$~$\ref{2}(a) and fig.$~$\ref{2}(b), we present the squeezing degrees $S_c$ , $S_b$  for the microwave field, mechanical mode , respectively. Experimentally feasible parameters are used in the study~\cite{19,60,61}: $\omega_m/2\pi=10 $MHz, $\omega_c/2\pi=10 $GHz, $\omega_a/2\pi=200$THz, mechanical quality factor $Q=5\times10^4$, $\kappa_a=0.08\omega_m$, $\kappa_c = 0.005\omega_m$, $T=15$mK, $G_- = \kappa_a$ (the corresponding optical driving power $P_-\approx1.35$mW for frequency $\omega_-=\omega_a-\omega_m$), $G_c=0\sim1\kappa_a$ (the range of microwave driving power $P_c$ is $0\sim79.40$mW). All subsequent analyses are performed within the stable regime. It can be seen from fig.$~$\ref{2}(a) and fig.$~$\ref{2}(b) that the squeezing degree of the two modes will change significantly with the variation of the ratio $G_+/G_-$. And the $S_c$ and $S_b$ achieve the value of strong squeezing when $G_+/G_-\approx0.78$ and $G_c=0.6\kappa_a$, which are 4.61 and 4.56 respectively. This can be explained by Eq.$~$(\ref{x18}): on the one hand, in order to achieve a stronger coupling between optical mode and mechanical mode so as to more effectively cool the mechanical mode to the quantum ground state, a sufficiently large gap between $G_+$ and $G_-$ is required to obtain a larger $G'$. On the other hand, if the gap between $G_+$ and $G_-$ is too large, it will lead to a reduction in the interaction between the mechanical mode and the microwave mode, which is not conducive to the transfer of the squeezing. Therefore, there exists an optimal value for $G_+/G_-$, which enables the squeezing of the mechanical mode and the microwave mode to reach the relative maximum value, this optimal ratio is close to 1. When $G_+/G_- = 0.78$, $G_c = 0.8G_-$, we presente the Wigner functions of the microwave mode and the mechanical mode in fig.$~$\ref{2}$(c)$ and fig.$~$\ref{2}$(d)$, from which it can be clearly seen that the orthogonal components of both modes have been simultaneously squeezed. And they are the phase component of the microwave mode and the displacement component of the mechanical mode.
\begin{figure}
	\centering
	\includegraphics[height=3.5cm,width=4.2cm]{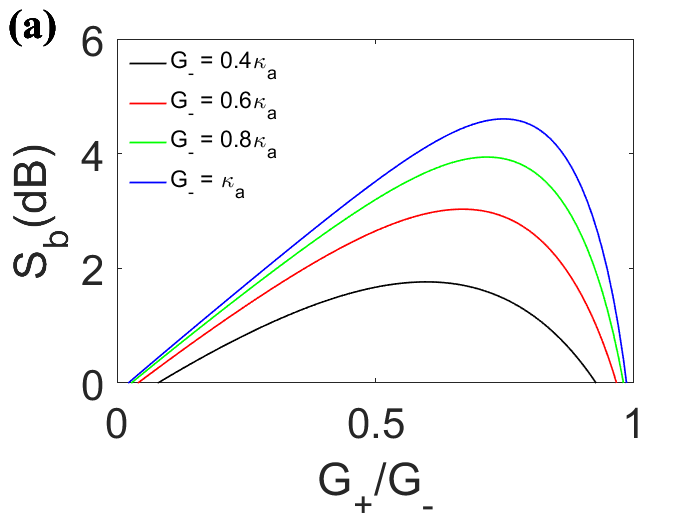}
	\includegraphics[height=3.5cm,width=4.2cm]{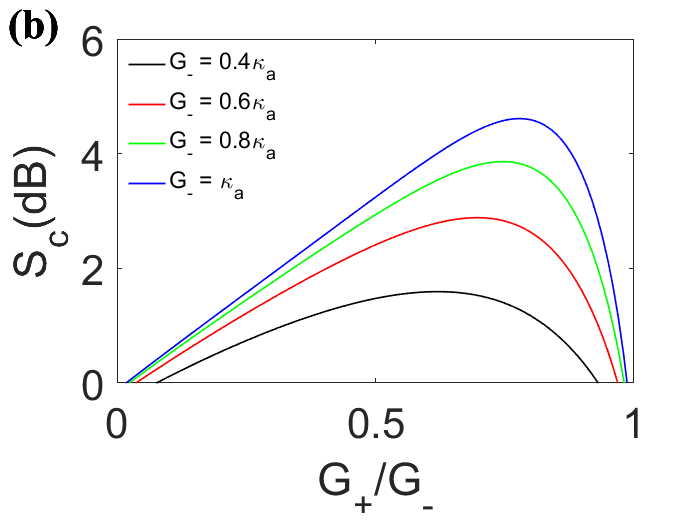}
	\caption{The squeezing degrees of $S_b$ (a) and $S_c$ (b) versus $G_+/G_-$ with fixed ratio $G_c/G_-=0.6$. The four lines correspond to $G_-=[0.4\kappa_a,~0.6\kappa_a,~0.8\kappa_a,~1.0\kappa_a]$ , respectively. The remaining parameters are the same as those in Fig.$~$\ref{2}.}
	\label{3}
\end{figure}

In fig.$~$\ref{3}, we show the squeezing degrees of $S_c$ and $S_b$ with fixed ratio $G_c/G_-=0.6$ when $G_-=[0.40\kappa_a,~0.60\kappa_a,~0.80\kappa_a,~1.00\kappa_a]$, $Gc=[0.24\kappa_a,~0.36\kappa_a,~0.48\kappa_a,~0.60\kappa_a]$ accordingly, and the corresponding $P_c \approx [4.58,~10.30,~18.30,~28.60]$(mW). By comparing the lines in  fig.$~$\ref{3}(a) and fig.$~$\ref{3}(b), it can be observed that as $G_-$ increases, the mechanical mode was further cooled to the quantum ground state, and its squeezing degree increased, and the effective coupling strength between the mechanical mode and the microwave mode correspondingly increases, representing the rate of energy exchange between the two modes has accordingly increased from Eq.$~$(\ref{x18}). Consequently, the squeezing of both modes also increases simultaneously, with the maximum value exceeding 4.5.

\begin{figure}
	\centering
	\includegraphics[height=3.5cm,width=4.2cm]{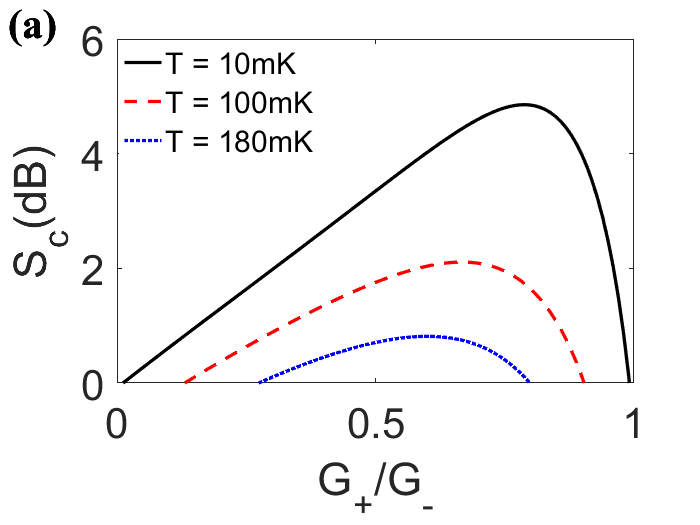}
	\includegraphics[height=3.5cm,width=4.2cm]{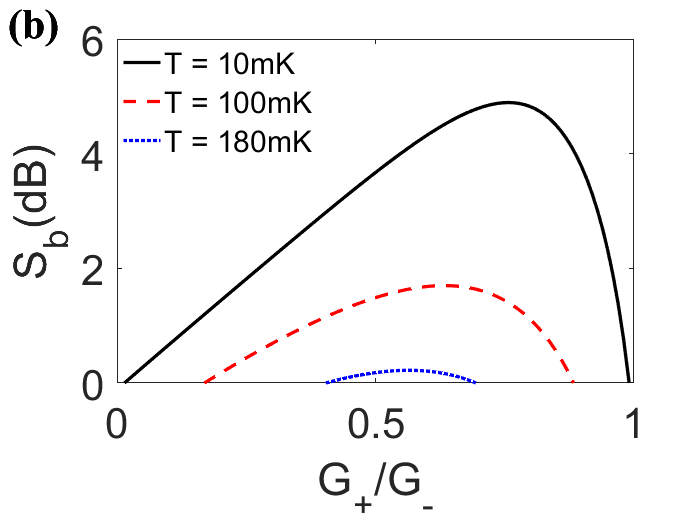}
	\caption{The squeezing degree of the microwave mode ($\hat{a}$) and the mechanical mode ($\hat{b}$) at different temperatures with $G_-=\kappa_a$ and $G_c=0.6G_-$. The three selected temperatures are $T=[10,~100,~180]$(mK). The remaining parameters are the same as those in Fig.$~$\ref{2}. }
	\label{4}
\end{figure}

Now, based on the analysis in the previous text, we take $G_-=\kappa_a$ and analyze the variations of the two squeezing degrees with temperature in fig.$~$\ref{4}. We select three temperature gradients, namely 10mK, 100mK and 180mK. As the temperature rises, the environmental thermal noise of the system correspondingly increases, which disrupts the quantum effects between subsystems and consequently causes the quantum squeezing of the MC and the MR to gradually weaken. For this reason, we need to increase $G'$ to ensure the energy exchange between the mechanical mode and the optical mode is accelerated. Since $G'=\sqrt{G_-^2-G_+^2}$, when $G_-$ takes a fixed value $\kappa_a$, a larger $G'$ corresponds to a smaller $G_+$, which is equivalent to a smaller ratio of $G_+/G_-$. Therefore, as the temperature rises, the ratio of $G_+/G_-$ at the maximum squeezing degree of the two modes gradually decreases. In conclusion, we find that the squeezing of the mechanical mode and the microwave mode persists even at an ambient temperature as high as $0.18K$, which indicates that the squeezing of these two modes are robust against environmental thermal noise.

\section{Steady-state entanglement and quantum EPR steering}\label{Sec.IV}

Subsequently, we investigate the quantum entanglement and steering between the optical cavity and the mechanical oscillator. The quantum properties of these two subsystems can be seen from their corresponding sub-block in the covatiance matrix. Therefore, we exract and retain only the rows and columns related to these two modes, forming a 4×4 matrix:
\begin{align}
	\sigma_{ab}=\left[ \begin{array}{cccccc}
		V_a&&V_{ab}\\
		V_{ab}^T&&V_b
	\end{array} \right ],
\end{align}\label{x19}
where $V_a$, $V_b$, $V_{ab}$ are all 2×2 matrices. We employ logarithmic negativity to quantify the entanglement between optical mode and mechanical mode~\cite{2}:
\begin{align}
	EN_{ab}=\mathrm{max}[0,-\mathrm{ln}(2\eta^-)],
	\label{x20}
\end{align}
here $\eta^-\equiv2^{-1/2}\left\{\Sigma-[\Sigma^2-4\mathrm{det}\sigma_{ab}]^{1/2}\right\}^{1/2}$ is the smallest symplectic eigenvalue of matrix $\sigma_{ab}$, and $\Sigma\equiv\mathrm{det}V_a+\mathrm{det} V_b-2 \mathrm{det} V_{ab}$. It can be seen from Eq.$~$(\ref{x20}) that when $\eta^-<0.5$, there exists entanglement between the OC and the MR. We calculate EPR steering in the manner described in Ref.~\cite{59}, the expressions for the conversion from optical mode to mechanical mode and the reverse conversion are:
\begin{align}\label{21}
	ST_{ab} &= \max \left\{ 0, \, S(2V_a) - S(2\sigma_{ab}) \right\},  \notag\\
	ST_{ba} &= \max \left\{ 0, \, S(2V_b) - S(2\sigma_{ab}) \right\}, 
\end{align}
here the Rényi-2 entropy $S(\sigma)=\frac{1}{2}$ln det$\sigma$. The values of $ST_{ab}$ and $ST_{ba}$ quantify the degree of steering. When $ST_{ab}$ or $ST_{ba}$ is greater than zero, there exists steering in the corresponding direction. 
\begin{figure}
	\centering
	\includegraphics[height=3.5cm,width=4.2cm]{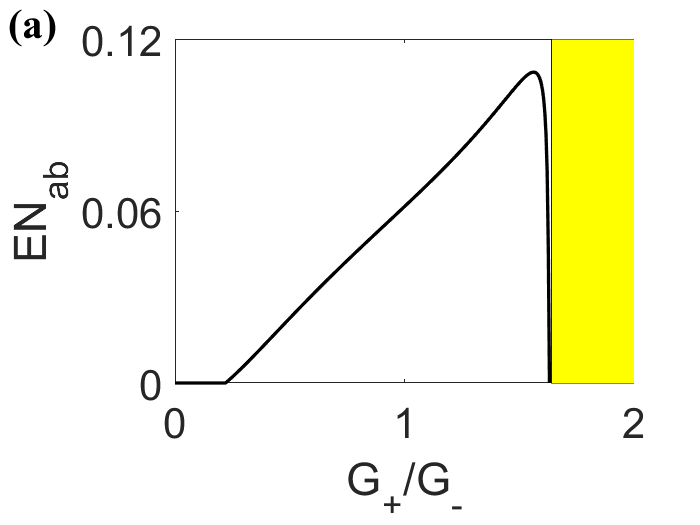}
	\includegraphics[height=3.5cm,width=4.2cm]{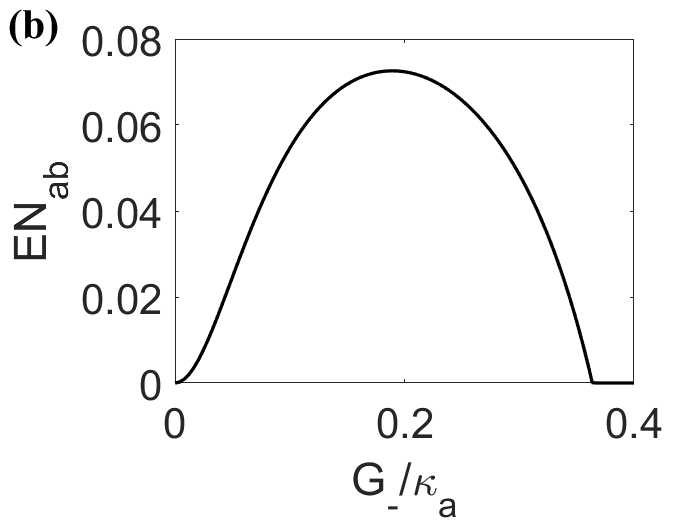}
	\caption{The steady-state quantum entanglement between the OC and the MR versus the ratio $G_+/G_-$ (a) with $G_-=~0.2\kappa_a$ , corresponding to $P_- \approx 0.06$(mW), and $G_-/\kappa_a$ (b). The yellow area in (a) represents the unstable region. In (b), $G_-=0\sim0.4\kappa_a$, $G_+=1.15G_-$, and $G_c=0.4\kappa_a$. The remaining parameters are the same as those in Fig.$~$\ref{2}.}
	\label{5}
\end{figure}
\begin{figure}
	\centering
	\includegraphics[height=3.5cm,width=4.2cm]{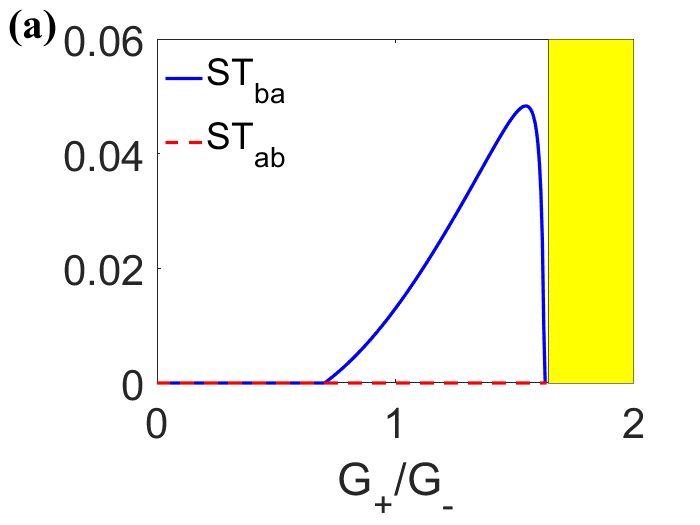}
	\includegraphics[height=3.5cm,width=4.2cm]{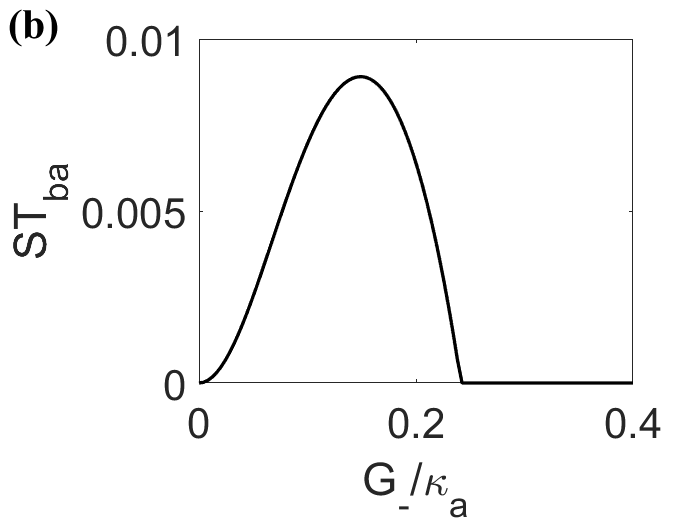}
	\caption{(a) The steady-state EPR steering between the OC and the MR versus the ratio $G_+/G_-$ with $G_-=0.2\kappa_a$, corresponding to $P_- \approx 0.06$(mW), $G_c = 0.4G_-$, the yellow area represents the unstable region, (b) $ST_{ba}$ versus $G_-/\kappa_a$ with $G_-=0\sim0.4\kappa_a$, $G_+=1.15G_-$ and $G_c=0.4\kappa_a$. The remaining parameters are the same as those in Fig. 2.}
	\label{6}
\end{figure}

The steady-state entanglement between the OC and the MR with respect to the ratio $G_+/G_-$ is shown in Fig.$~$\ref{5}. In Fig.$~$\ref{5}(a), optical driving power $P_-=$ 0.06 mW and
$G_c/G_-=0.4$ are adopted. We find that as the optical driving power $P_-$ increases, the two-mode entanglement $EN_{ab}$ initially increases and then decreases, a more clearly trend is shown in Fig.$~$\ref{5}(b) when $G_+/G_-$ is held constant. This can be interpreted as follows: as $P_-$ increases, under the condition of fixed $G_+/G_-~=~1.15$, the effective coupling strength of the parametric amplification interaction between $\hat{a}$ and $\hat{b}$ increases. However, when $P_-$ reaches the optimal value for achieving maximum entanglement, further increases in $P_-$ lead to intensified thermal noise within the system, which disrupts the quantum effects and reduces the degree of entanglement. Besides, due to the small value of $G_c$, under the selected parameters, the entanglement $EN_{bc}$ of the mechanical mode and the microwave mode is much smaller than that of the optical mode and the mechanical mode ($EN_{ab}$). Therefore, $EN_{bc}$ is not considered here. It is worth noting that the driving laser power here is two orders of magnitude lower than that in Ref.~\cite{19}, however, the presented scheme can still achieve the same order of magnitude of entanglement between the optical mode and the mechanical mode as described in Ref.~\cite{19}. This is because the dual-light driving introduces the two-mode squeezing interaction between the two modes into the effective Hamiltonian Eq.$~$(\ref{x9}).

\begin{figure}
	\centering
	\includegraphics[height=5.5cm,width=6.2cm]{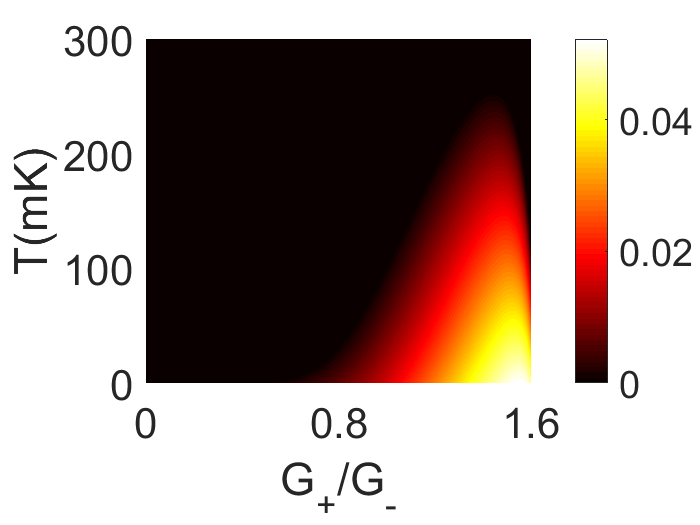}
	\caption{ This picture represent steady-state EPR steering between the OC and the MR with $G_-=~0.2\kappa_a$, $T = 0\sim300$mK, $G_c = 0.4G_-$, the remaining parameters are the same as those in Fig.$~$\ref{2}. }
	\label{7}
\end{figure}

Using the experimentally feasible parameters identical to those in Fig.$~$\ref{5}, the EPR steering between the optical mode and the mechanical mode is illustrated in Fig.$~$\ref{6}. It is obvious that the pairs of $ST_{ab}$ and $ST_{ba}$ correspond to the subsets of $EN_{ab}$. In the subsystem composed of the OC and the MR, the dissipation of the optical mode is much greater than that of the mechanical mode. Moreover, by applying a strong driving microwave cavity, the displacement of the mechanical mode will be amplified through the coupling term $\hbar{g}_{c}\hat{c}^\dagger\hat{c}(\hat{b}^\dagger+\hat{b})$ (can also be rewritten as $\hbar{g}_{c}{|c_s|}^2(\hat{b}^\dagger+\hat{b})$), which further enhances the coupling efficiency of the mechanical mode and the optical mode. Such an asymmetric subsystem will generate significant asymmetric quantum steering. As shown in Fig.$~$\ref{6}, the numerical results indicate a perfect one-way steering between them. The steering value first increases with the rise of driving optical and microwave power. Subsequently, as the driving laser power is further increased, the quantum effects between subsystems are disrupted by thermal noise, leading to a decrease in the steering value. This trend of first increasing and then decreasing is further demonstrated in Fig.$~$\ref{6}(b). Further more, in Fig.$~$\ref{7}, we present the values of one-way steering at different temperatures. As can be seen from the figure, there is still one-way steering even at a temperature as high as 255mK.

\section{Conclusions}\label{Sec.V}

In conclusion, based on an electro-optomechanical system composed of an OC, an LC microwave circuits and an MR, we have proposed methods to achieve strong squeezing of microwave mode and mechanical mode as well as perfect one-way steering between optical mode and mechanical mode. The stability region, and the steady-state quantum fluctuations of the system can be obtained by solving the linearized dynamical equations of the system. With achievable experimental parameters, the squeezed state generated by cooling the MR with the dual-light driving can be transferred to the MC. We found that by adjusting the driving power of the system, the squeezing degree of the microwave mode could reach as high as 4.27. Meanwhile, the maximal squeezing degree of the microwave mode is 4.57. This property is of great significance for quantum information processing. Furthermore, we have also achieved the one-way EPR steering between the optical mode and the mechanical mode with weaker driving laser power compare to previous scheme~\cite{19}. Moreover, the degree of steering can be adjusted by the driving optical power, which will further expand the application of electro-optomechanical system in quantum secure communication.

\section*{Acknowledgments}
This work is supported by the National Natural Science Foundation of China under Grant No. 12274274.
\setlength{\parskip}{0pt}
\vspace{-0.6cm}
\appendix

\end{document}